\documentclass[twocolumn,floatfix, showpacs]{revtex4}

\usepackage[final]{epsfig}
\usepackage{amsmath}
\usepackage{parskip}
 
\def\lsim{\,\raise0.3ex\hbox{$<$\kern-0.75em\raise-1.1ex\hbox{$\sim$}}\,}
\def\gsim{\,\raise0.3ex\hbox{$>$\kern-0.75em\raise-1.1ex\hbox{$\sim$}}\,}
 
\begin{document}
 
\title{A Monte-Carlo model for elastic energy loss in a hydrodynamical background}
 
\author{Jussi Auvinen}
\email{jussi.a.m.auvinen@jyu.fi}
\author{Kari J.~Eskola}
\email{kari.eskola@phys.jyu.fi}
\author{Thorsten Renk}
\email{trenk@phys.jyu.fi}
\affiliation{Department of Physics, P.O. Box 35, FI-40014 University of Jyv\"askyl\"a, Finland}
\affiliation{Helsinki Institute of Physics, P.O. Box 64, FI-00014, University of Helsinki, Finland}
 

\begin{abstract}
We present a computation of elastic energy loss of hard partons traversing the bulk hydrodynamical medium created in ultrarelativistic heavy-ion collisions. The model is based on perturbative Quantum Chromodynamics (pQCD) cross sections for $2\rightarrow 2$ processes in which a hard incoming parton is assumed to interact with a thermal parton from the medium. We model the interactions of this type in a Monte-Carlo framework to account properly for exact energy-momentum conservation, non-eikonal parton propagation, parton conversion reactions and the possibility to create additional hard recoiling partons from the medium. For the thermodynamical properties of the medium we use a hydrodynamical evolution model. We do not aim at a full description of high transverse momentum ($P_T$) observables at this point. Rather, we view the model as a starting point in obtaining a baseline of what to expect under the assumptions that 
the medium is describable by thermal quasifree partons and that their pQCD interactions with the high-energy partons are independent. Deviations from this baseline then call for more sophisticated medium description, as well as inclusion of higher-order processes and coherence effects in the pQCD scatterings.
\end{abstract}

\maketitle

\section{Introduction}

The energy loss of hard partons propagating through the soft medium created in heavy-ion collisions has long been regarded as a promising tool to gain information on the medium properties \cite{Jet1,Jet2,Jet3,Jet4,Jet5,Jet6}. Radiative energy loss, i.e. the idea that medium-induced radiation predominantly carries away energy from a hard parent parton has been rather successful in describing not only the hadronic nuclear suppression factor for central collisions but also the effects of changing medium geometry \cite{Dainese}. Calculations within dynamical evolution models in various formalisms \cite{Dyn01,Dyn02,Dyn03} have improved on this result and show also agreement with measured hard back-to-back correlations \cite{Correlations1,Correlations2,Correlations3} and the measured suppression of protons \cite{ppbar}.

Nevertheless, there are indications that a radiative energy loss picture fails to describe the suppression of heavy quarks as seen in the measurements of the single-electron spectra \cite{HQPuzzle}. Energy loss due to elastic collisions with the medium \cite{Thoma1,Thoma2,Thoma3,Mustafa,Mustafa2,DuttMazumder,Djordjevic,Wicks,Ruppert} has been suggested as a possible solution to this problem. Such calculations indicate a large component of elastic energy loss also for light quarks and gluons. However, the pathlength dependence of elastic energy loss does not agree with the observed response of the suppression to a large pathlength bias as probed in back-to-back hadron correlations \cite{ElasticPhenomenology}.

In the light of these inconclusive findings, it would be desirable to have a baseline calculation of elastic energy loss which treats the relevant physics in as detailed manner as possible. In particular, many analytical calculations of elastic energy loss assume eikonal propagation of a hard parton, however in a more realistic description even hard partons, $p_T\gsim 5$~GeV, can be deflected from their path by interactions with the medium. Similarly, in processes like $gg\rightarrow q\overline{q}$ there is not only a change of energy between incoming and outgoing parton but also a change in parton species which influences the subsequent interaction probability of the outgoing parton. Finally, in scattering processes in which the energy is distributed almost evenly among the outgoing partons, it is not justified to use the picture of energy from a hard parton being lost into the medium. Clearly, one also has to account for situations in which both partons after a scattering process are hard or both become part of the medium, or in which the identity of the leading parton changes.

In the present manuscript, we present work towards such a baseline computation. For this purpose, we develop a Monte Carlo (MC) simulation for the hard parton's interaction with the medium, as this is an appropriate framework to account for all the above mentioned effects. In addition, we also try to to give estimates on the uncertainties of such a computation. Similar, in some aspects even more ambitious approaches are the MC models JEWEL (Jet Evolution With Energy Loss) \cite{JEWEL} and MARTINI (Modular Algorithm for Relativistic Treatment of heavy IoN Interactions) \cite{Schenke:2009gb}, both of which consider pQCD interactions of a parton shower with a thermal medium and include both elastic and radiative energy-loss components. 
In JEWEL, one models shower evolution and medium-induced radiation in a static (no flow, $T$ constant) medium, and in MARTINI the vacuum showering is assumed to cease when the hydrodynamic evolution starts, while our approach focuses on the leading partons. Like MARTINI, also   
we utilize a realistic well-tested 3-dimensional hydrodynamical description of the medium (in our case azimuthal symmetry in central collisions, and longitudinal boost symmetry but 3-d flow) instead of a static medium applied in JEWEL.

\section{The model}

In the following, we make the assumption that the medium can be characterized as a thermal gas of quasi-free quarks and gluons and that its interactions with a hard parton can be treated as incoherent $2\rightarrow2$ processes in leading-order pQCD. We stress that these assumptions need not be realized for the medium produced in ultrarelativistic heavy-ion collisions --- any discrepancy between the calculation presented here and the data will reflect this.

We model the medium by means of a hydrodynamical calculation \cite{Hydro} with the initial state obtained from the saturation of the phase space with minijets \cite{EKRT}. This provides the medium temperature $T$ at each spacetime point. For the moment, we assume that only the scatterings with a partonic medium are relevant and ignore the possibility that a hadronic medium might interact with a hard parton. Thus the mean free path of hard partons becomes infinite once the medium temperature drops below the critical temperature $T_C$. Given the local $T$, the distribution of scattering partners is given by the Fermi-Dirac (Bose-Einstein) distribution for quarks (gluons) respectively. We 
also account for the 3-d hydrodynamical flow of the fluid cell.

Our approach, most specifically the assumption that the relevant interactions can be described in terms of partonic pQCD interaction, shares many features with parton cascade models (e.g. \cite{PCM1,PCM2,PCM3}). However, unlike in parton cascade models where the partonic pQCD interactions are assumed to describe the whole off-equilibrium bulk medium, we assume that the bulk is essentially thermalized and use explicit pQCD interactions only to account for the non-equilibrium dynamics between hard parton and medium.

In the following, we first give a more detailed account of the hydrodynamical simulation, the parton-medium scattering rates and details of the MC implementation. Then we present our results for the partonic and hadronic nuclear modification factor $R_{AA}(p_T)$ and for QCD-matter tomography, and also discuss the relevant uncertainties. Some technical details 
regarding the scattering rate calculations can be found in the Appendix. 

\subsection{Hydrodynamics}

In the present exploratory stage we consider only the cleanest possible case, (nearly) central $A+A$ collisions in the mid-rapidity region. The produced QCD matter, which evolves and flows 3-dimensionally, can to a first realistic approximation then be described by using the nonviscous, azimuthally symmetric, longitudinally boost-invariant hydrodynamic framework. The 1+1 d hydrodynamical framework, which we apply here for the QCD medium evolution, is presented in detail in Ref.~\cite{Hydro}. The initial energy density $\epsilon(r,\tau_0)=\epsilon_0(r)$ for the hydrodynamics is here obtained from pQCD-minijet production \cite{Eskola:1988yh} and final-state saturation as computed in the EKRT-model \cite{Eskola:1999fc}. The transverse profile of $\epsilon_0$ is assumed to scale with the number of binary collisions, and thermalization is assumed to take place at formation (i.e. at saturation). For central Au+Au collisions at $\sqrt s_{NN}=200$~GeV, we thus initialize the hydrodynamics at (see Table 1 in \cite{Hydro}) $\tau=\tau_0=0.17$~fm/$c$, where the longitudinal proper time is defined as $\tau \equiv \sqrt{t^2-z^2}$. The transverse flow is assumed to be zero at $\tau_0$. The Equation of State (EoS) in our hydrodynamic background contains a high-$T$ QGP phase, a low-$T$ hadron resonance gas (HRG) phase, and a mixed phase between them. The QGP here is an ideal gas of gluons and three flavours of massless quarks and antiquarks obeying a Bag-EoS, while the HRG is an ideal gas of all hadrons and hadron resonances up to $M=2$~GeV. The Maxwell construction gives the mixed phase, and the Bag constant $B^{1/4} =239$~MeV fixes the phase transition temperature to $T_c = 165$~MeV. We terminate the hydrodynamic evolution at a single decoupling temperature $T_{\rm dec}= 150$~MeV. As discussed in Ref. \cite{Hydro}, the measured pion and kaon spectra and proton multiplicity in central Au+Au collisions at RHIC can be reproduced very nicely with such hydrodynamical set-up, once the thermal hadron spectra have been computed via the Cooper-Frye decoupling procedure and once the strong and electromagnetic hadronic decays have been accounted for. For more details, as well as for plots of the initial energy density and of the phase boundary locations in the $(\tau,r)$ plane, we refer the reader to Ref. \cite{Hydro}.

\subsection{Scattering rate and mean free path}
\label{subsec:rates}

Our simulation of energy losses of high-energy partons in the produced QCD matter rests largely on one particular physical quantity, namely the scattering rate for a high-energy parton of a type $i$, 
\begin{equation}
\label{totgamma}
\Gamma_i (p_1,u(x),T(x)) = \sum_{j(kl)} \, \Gamma_{ij\rightarrow kl}(p_1,u(x),T(x)),
\end{equation}
where all possible elastic and inelastic partonic processes $ij\rightarrow kl$ are accounted for by summing over
all types of collision partners, $j=u,d,s,\bar u, \bar d, \bar s,g$ in the initial state, and over all possible parton type pairs $(kl)$ in the final state. 
In general, the scattering rate depends on the frame, and in particular on the high-energy parton's 4-momentum $p_1$, on the flow 4-velocity $u(x)$ and on the temperature $T(x)$ of the fluid at each space-time location $x$. In the notation below, we leave these dependences implicit but in the simulation they are fully accounted for. Ignoring Pauli blocking and Bose enhancement, we express the scattering rate as follows (see also Ref. \cite{Matsui})

\begin{equation}
\label{scattrate}
\begin{split}
\Gamma_{ij\rightarrow kl} = &\frac{1}{2E_1} \int \negthickspace \frac{d^3p_2}{(2\pi)^3 2E_2} \negthickspace \int \negthickspace \frac{d^3p_3}{(2\pi)^3 2E_3} \int \negthickspace \frac{d^3p_4}{(2\pi)^3 2E_4} \\
& f_j(p_2\cdot u,T) \, |M|_{ij\rightarrow kl}^2(s,t,u) \, S_2(s, t, u) \\
& (2\pi)^4 \delta^{(4)}(p_1+p_2-p_3-p_4).
\end{split}
\end{equation} 

where \(|M|^2_{ij\rightarrow kl}\) is the spin- and colour-summed/averaged scattering amplitude which depends on the standard Mandelstam variables $s$, $t$ and $u$.  The distribution function for a thermal particle of momentum \(p_2\) is denoted by \(f_j(p_2\cdot u,T)\), which is the Bose-Einstein distribution $f_g = f_B(p_2\cdot u,T)$ for gluons and the Fermi-Dirac distribution $f_q = f_D(p_2\cdot u,T)$ for quarks. Since the initial state spins and colours have been averaged over in the squared matrix elements, the spin- and color-degeneracy factors in $f_B$ and $f_D$ are the standard ones, $g_g=2\cdot8=16$ for gluons, and $g_q=2\cdot3 = 6$ for quarks and antiquarks, while the possible flavour-degeneracy needs to be determined separately in each process. In our simulation, we evaluate the scattering rate always in the local rest-frame of the fluid, so that $p_2\cdot u=E_2$ and $E_1$ is the energy of the high-energy parton $i$ in this frame. 

The singularities arising in forward and backward scatterings of massless partons --- appearing in the scattering amplitudes at $t,u\rightarrow 0$ --- need to be regularized in some physically meaningful way. 
For the moment, an exact field-theoretical solution to this non-equilibrium problem is not known.
One possible treatment is to apply resummed thermal (equilibrium) propagators separately for the longitudinal and transverse components, as is done in the AMY approach \cite{Arnold:2002ja,Ruppert}. Since the problem is, however, still phenomenological, we choose here a more transparent approach (tested also in JEWEL before) and introduce a thermal-mass-like overall cut-off scale $m=s_mg_sT$ for the momentum phase space, such that $u,t\le -m^2$. Here $g_s$ is the strong coupling constant and $s_m$ is a parameter of the order of one but which we can vary.
As we cannot control the running of $\alpha_s = \frac{g_s^2}{4\pi}$, we keep the strong coupling constant fixed with momentum scale but vary both $g_s$ and $s_m$ in different runs of the MC code to test the model sensitivity to these parameters.
Since $s+u+t=0$, the cut-off conditions are equivalent to $-s+m^2\le t\le -m^2$ with the requirement $s\ge 2m^2$. These are encoded in the function $S_2$ above, which is defined as
\begin{equation}
\label{cutoff}
S_2(s, t, u) = \theta(s\ge 2m^2)\theta(-s+m^2\le t\le -m^2)
\end{equation}

For the evaluation of the Lorentz-invariant part in Eq.~\eqref{scattrate}, it is useful to recall that the scattering rate can be expressed in terms of the scattering cross sections. For the process $ij\rightarrow kl$, where the Lorentz-invariant cut-off conditions are imposed, and the identities of the final state particles are accounted for, the cross section is
\begin{equation}
\label{gencrosssection}
\begin{split}
\sigma_{ij\rightarrow kl}(s) =& \frac{1}{1+\delta_{kl}}\frac{1}{2s} \int \, \frac{d^3p_3}{(2\pi)^3 2E_3} \int \, \frac{d^3p_4}{(2\pi)^3 2E_4}
S_2(s,t,u) \\& \cdot (2\pi)^4 \delta^{(4)} (p_1+p_2-p_3-p_4) \, |M|_{ij\rightarrow kl}^2(s,t,u) ,\\
=&  \frac{1}{1+\delta_{kl}}\frac{1}{16\pi s^2} \int_{-s+m^2}^{-m^2} dt \, |M|_{ij\rightarrow kl}^2,
\end{split}
\end{equation}
and we can rewrite Eq.~\eqref{scattrate} as
\begin{equation}
\label{nsgamma}
\begin{split}
\Gamma_{ij\rightarrow kl} &= \int \, \frac{d^3p_2}{(2\pi)^3} f_j(E_2,T) \theta(s\ge 2m^2)
\frac{2s}{2E_1 2E_2} \sigma_{ij\rightarrow kl}(s) \\
&\equiv  n_j(T) \langle (1-\cos \theta_{12}) \sigma_{ij\rightarrow kl}(s\ge 2m^2)\rangle. 
\end{split}
\end{equation}
where \(n_j(T)\) is the thermal particle density in the rest frame of the QCD fluid, 
and where we have expressed the Mandelstam variable as $s=2E_1E_2(1-\cos\theta_{12})$, i.e. in terms of the angle \(\theta_{12}\) between the hard parton and the thermal parton, and their energies in the fluid rest frame.
Changing the integration variable $\cos\theta_{12}$ to $s$, it is straightforward to arrive at the following  result
\begin{equation}
\label{nsgamma_analytic}
\Gamma_{ij\rightarrow kl} = \frac{1}{16\pi^2E_1^2}\int_{\frac{m^2}{2E_1}}^{\infty}dE_2f_j(E_2) \Omega_{ij\rightarrow kl}(E_1,E_2,m^2),
\end{equation}
where 
\begin{equation}
\label{omegafunction}
\Omega_{ij\rightarrow kl}(E_1,E_2,m^2)=\int_{2m^2}^{4E_1E_2}ds [s\sigma_{ij\rightarrow kl}(s)].
\end{equation}
The analytical results for the relevant elements of $\Omega_{ij\rightarrow kl}$ are listed in the Appendix. The remaining $E_2$ integral is easiest to do numerically. 

\begin{figure}[t]
\centering
\includegraphics[width=9cm]{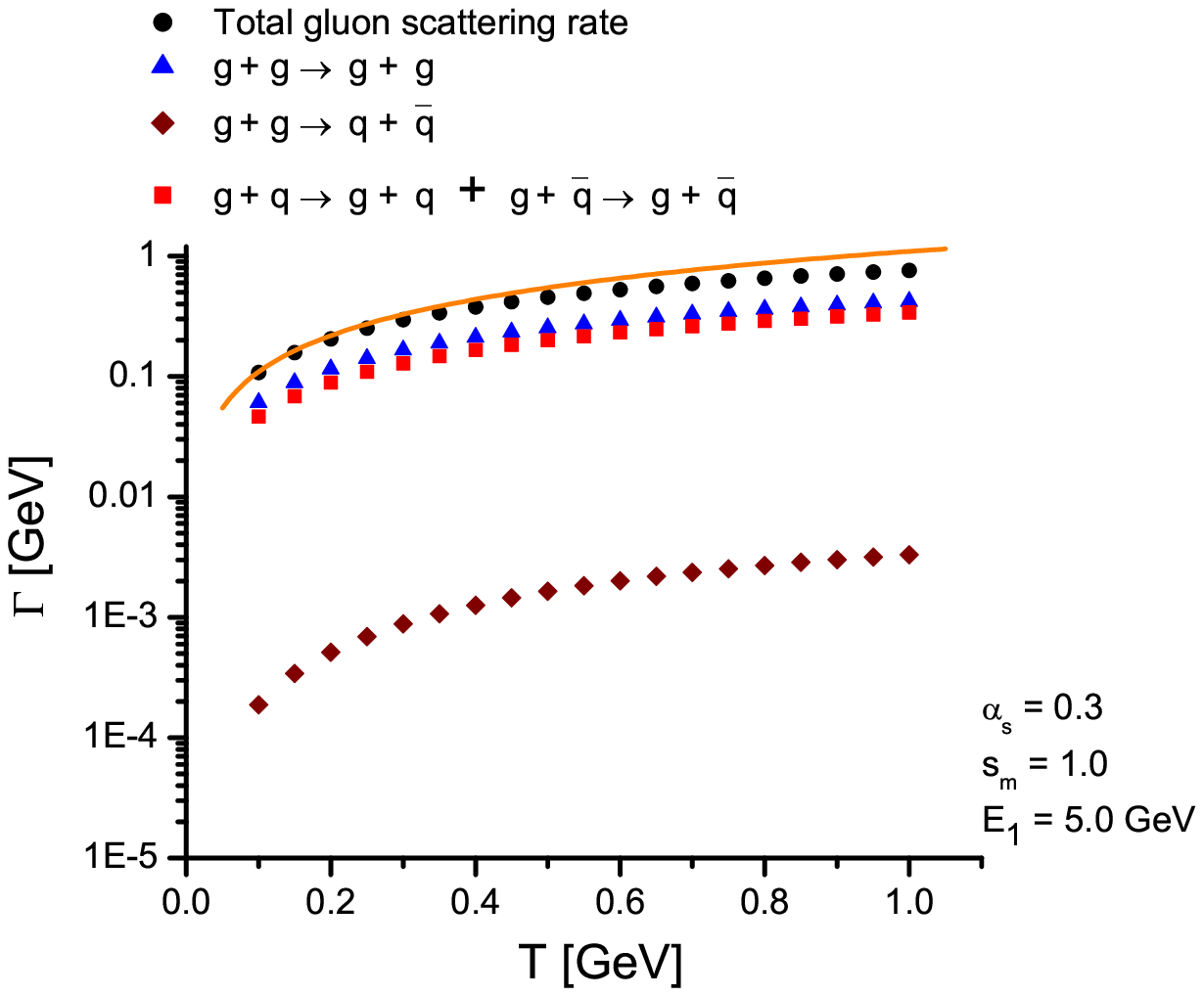}
\includegraphics[width=9cm]{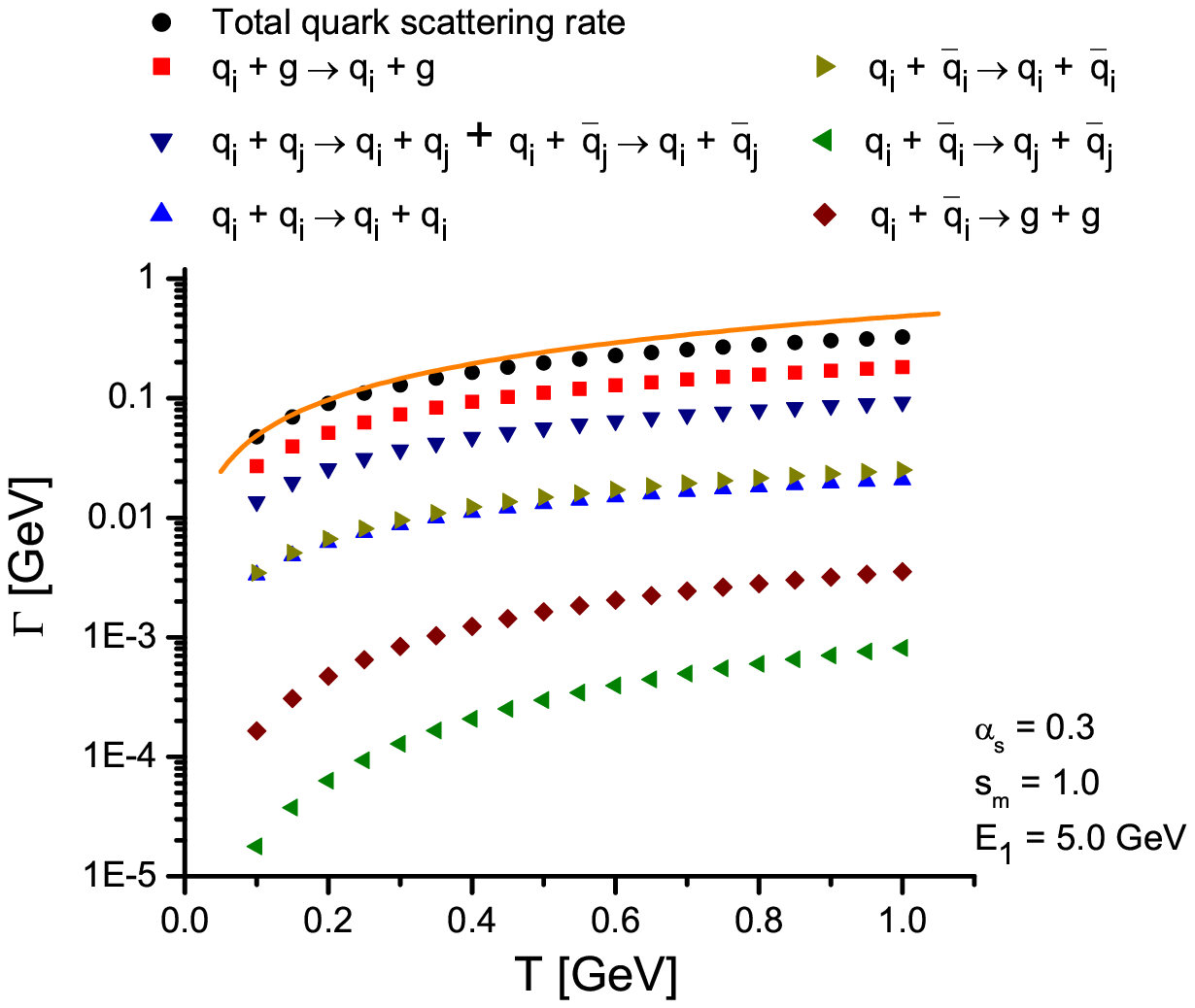}
\caption{(Color online) The scattering rates \(\Gamma_i\) of gluon (above) and quark (below) as a function of temperature \(T\)
for a QCD plasma at rest. Flavour- and quark-antiquark -summed contributions from different processes and the analytical estimates for the total rates discussed in the text are also shown (solid lines).}
\label{fig: rates_vs_T}
\end{figure}

\begin{figure}[t]
\centering
\includegraphics[width=9.0cm]{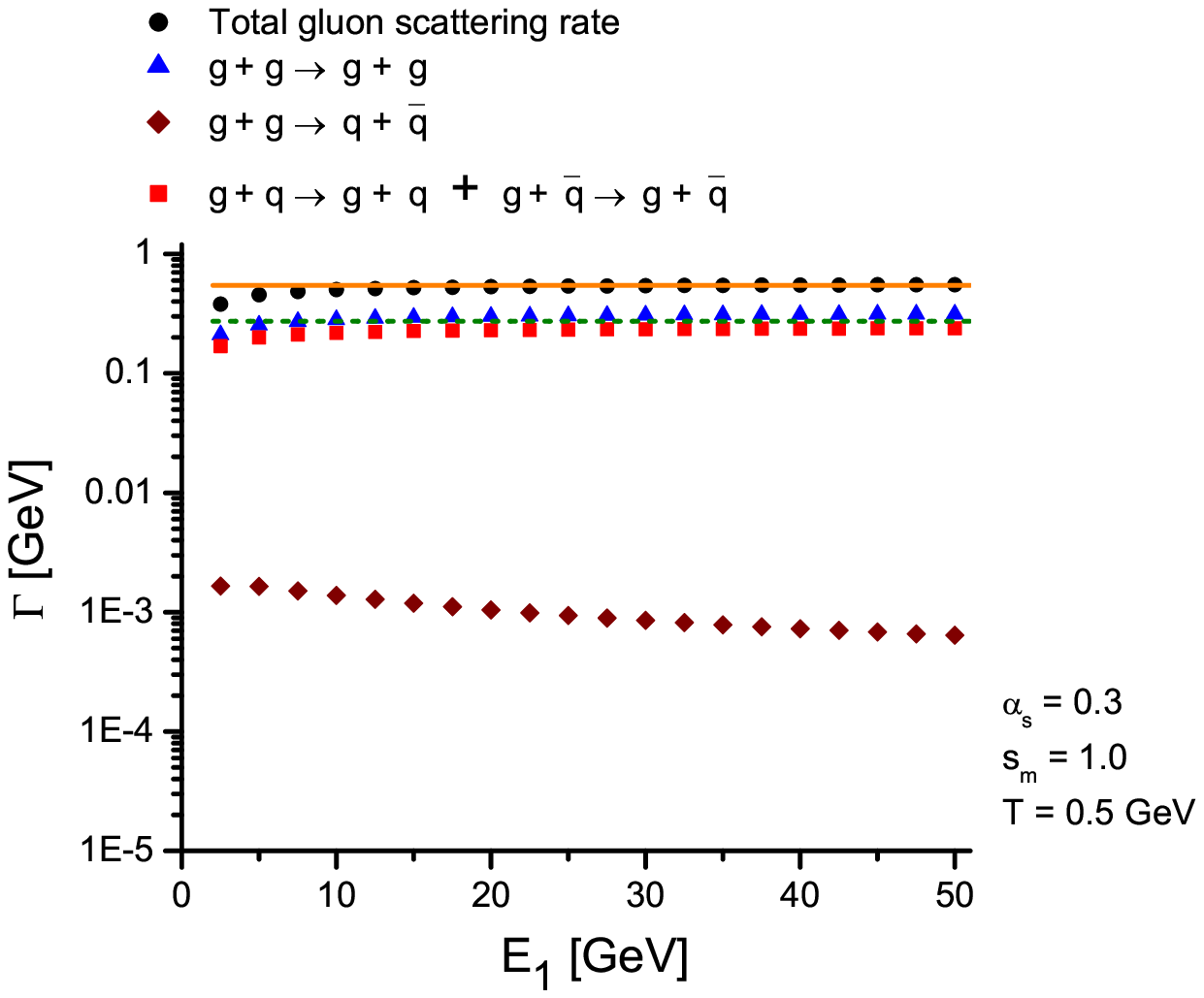}

\vspace{-0.2cm}
\includegraphics[width=9.0cm]{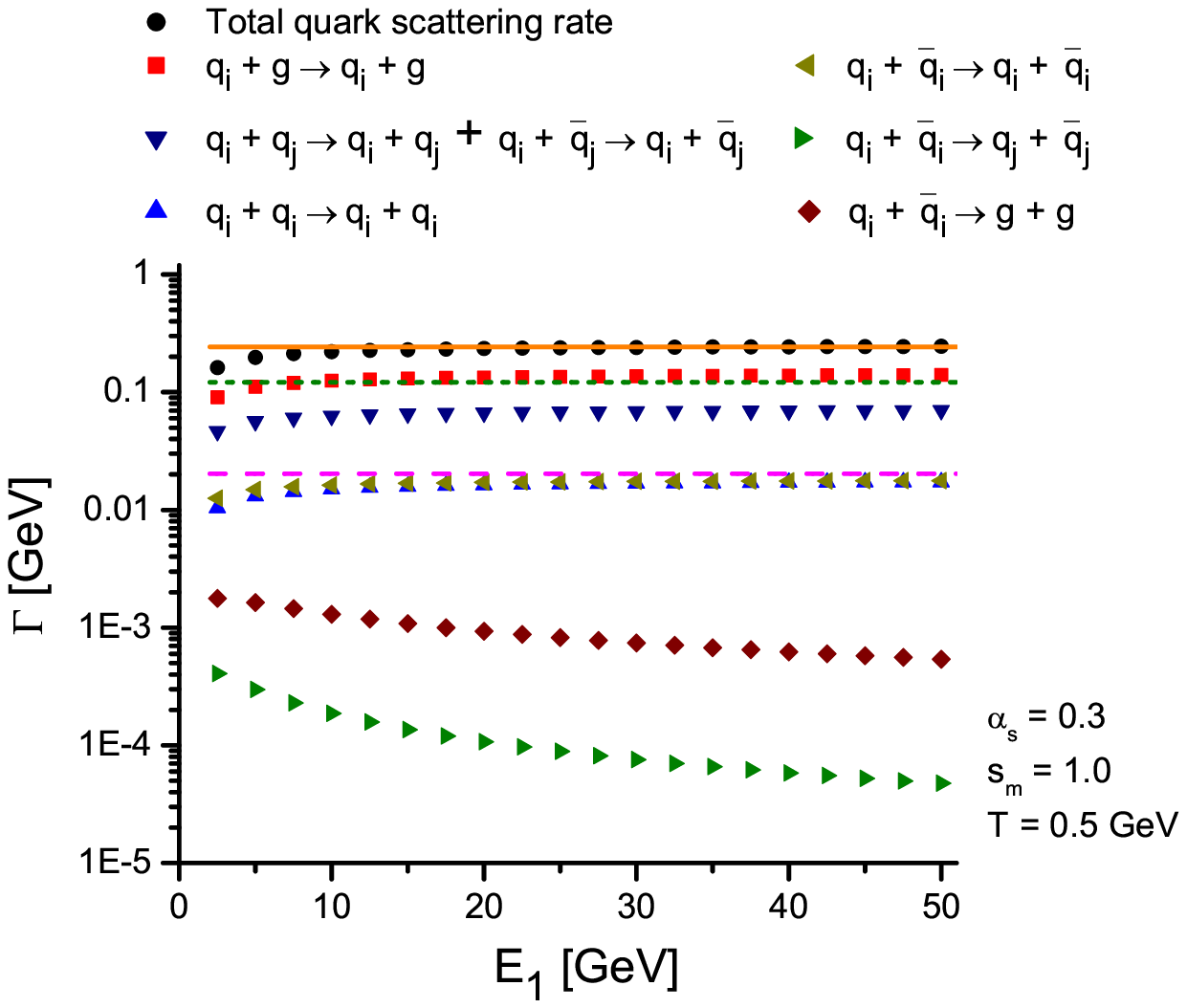}

\vspace{-0.2cm}
\caption{(Color online) The scattering rates \(\Gamma_i\) of gluon (above) and quark (below) as a function of the parton energy \(E_1\), for a QCD plasma at rest at $T=0.5$ GeV. Flavour-summed contributions from different processes  are shown. 
The analytical estimates for \(\Gamma_i\) (see text) are shown by the solid lines. In the upper panel,  the analytical estimate for the $gg\rightarrow gg$ process (dashed line) is the same as for the quark-antiquark summed $gq\rightarrow gq$ process.  In the lower panel, the dashed lines show the contributions from the $q_ig\rightarrow q_ig$ process and an individual $q_iq_j\rightarrow q_iq_j$ process. }
\label{fig: rates_vs_E1}
\end{figure}

The obtained scattering rates $\Gamma_i$ for quarks, antiquarks and gluons, along with the contributions $\Gamma_{ij\rightarrow kl}$ from different processes, are shown in Fig.~\ref{fig: rates_vs_T} as a function of temperature $T$ and in Fig.~\ref{fig: rates_vs_E1} as a function of the hard parton energy $E_1$.

To get a feeling of the general characteristics of the system, we consider also the following analytical estimates.
In the process $ij\rightarrow kl$, the average of a quantity $A$ 
in the plasma at a temperature $T$ can be computed in the fluid rest frame as 
\begin{equation}
\begin{split}
\langle 
A
\rangle_{ij\rightarrow kl} \equiv& 
\frac{1}{\Gamma_{ij\rightarrow kl}}
\int  \frac{d^3p_2}{(2\pi)^3} f_j(E_2,T)\cdot A\cdot \\
&\theta(s\ge 2m^2)(1-\cos \theta_{12}) \sigma_{ij\rightarrow kl}(s).
\end{split}
\end{equation}

At the high-energy limit, $E_1^2,s\gg m^2,T^2$, the relevant pQCD cross sections all behave as $\sigma\sim m^{-2}$, and 
$\sigma_{gg\rightarrow gg}\approx\frac{9}{4}\sigma_{gq\rightarrow gq}\approx(\frac{9}{4})^2\sigma_{q_i q_j \rightarrow q_i q_j}$ (see Appendix, take the limit $E_1\gg m\sim E_2$, and $s\gg |t|,|u|$). In the Boltzmann approximation we thus get 
\begin{equation}
\label{averages}
\langle s \rangle \approx 8E_1T, \quad\quad
\langle E_2 \rangle \approx 3T, \quad\quad
\langle \cos\theta_{12} \rangle \approx -\frac{1}{3}.
\end{equation}
from which we see that, in this limit, 
$\langle s \rangle = 2E_1\langle E_2 \rangle(1-\langle \cos\theta_{12} \rangle$). 
The last result in Eq.~\eqref{averages} indicates that the collisions of the high-energy parton with the medium partons happen on the average in an angle $110^\circ$.

Using the same approximation, we can easily obtain analytical estimates also for the scattering rates, 
\begin{eqnarray}
\label{analytic_gammag}
\Gamma_g &\approx& 
\Gamma_{gg\rightarrow gg} + \sum_i [\Gamma_{gq_i\rightarrow gq_i} + \Gamma_{g\bar q_i\rightarrow g\bar q_i}] \nonumber \\
&\approx& (g_g + 2\cdot3\cdot g_q\cdot{\textstyle \frac{4}{9}}) \frac{9}{2}\frac{\alpha_s^2T^3}{\pi m^2} = 144\frac{\alpha_s^2T^3}{\pi m^2}\\
\label{analytic_gammaq}
\Gamma_{q_i} &\approx& 
\Gamma_{{q_i}g\rightarrow {q_i}g} +  
\sum_{j}[\Gamma_{{q_i}{q_j}\rightarrow {q_i}{q_j}} + \Gamma_{{q_i}{\bar q_j}\rightarrow {q_i}{\bar q_j}}]\nonumber\\
& \approx& (g_g\cdot{\textstyle \frac{4}{9}} + 2\cdot3 \cdot g_q\cdot({\textstyle \frac{4}{9}})^2)\frac{9}{2}\frac{\alpha_s^2T^3}{\pi m^2} = 
64 \frac{\alpha_s^2T^3}{\pi m^2}
\end{eqnarray}
where the factors $2\cdot 3$ account for quarks and antiquarks times the number of quark flavours. 
These estimates are shown in Figs.~\ref{fig: rates_vs_T} and Fig.~\ref{fig: rates_vs_E1}, together with the actual rates applied in our simulation (once the boost to the local rest frame has been taken into account). We observe that the analytical rates describe the numerical ones (both the total rate and the ones from the main processes) quite well in the kinematic and temperature range of RHIC. 

In a static homogeneous QGP, the average number of collisions experienced by the high-energy parton 
of an average path length $L$, would, in the high-energy limit, be $L\Gamma_i$. Taking $L=5$~fm and using Eqs.~\eqref{analytic_gammag}-\eqref{analytic_gammaq} with $\alpha_s=0.3$ and $s_m=1$, we arrive at the range 14--5 collisions for gluons and 6--2 collisions for quarks at $T=500-165$~MeV. This gives us a simple order-of-magnitude estimate of the expected number of elastic scatterings for a hard parton.

\subsection{The Monte Carlo simulation}

To initiate the hard massless parton of a type $i$ in each event, we sample the LO pQCD single-jet production spectrum 
(for more details, see \cite{Eskola:2002kv}) at $y_i=0$ and at $p_{Tmin}\leq p_T\leq \sqrt s/2$, 
\begin{equation}
\begin{split}
\frac{d\sigma}{dp_T^2 dy_i}^{\hspace{-0.4cm}AB\rightarrow i+X}\hspace{-0.6cm}
= \int dy_1 dy_2 \sum_{(kl)}
\frac{d\sigma}{dp_T^2dy_1dy_2}^{\hspace{-0.6cm}AB\rightarrow kl+X} \\
\left[ \delta_{ki} \delta(y_i-y_1) + \delta_{li}\delta(y_i-y_2) \right]
\frac{1}{1+\delta_{kl}},
\end{split}
\end{equation}
where 
\begin{equation}
\frac{d\sigma}{dp_T^2dy_1dy_2}^{\hspace{-0.6cm}AB\rightarrow kl+X}
\hspace{-0.6cm} = \sum_{ab}x_1 f_{a/A}(x_1,Q^2) x_2f_{b/B}(x_2,Q^2) \frac{d\hat\sigma}{d\hat
t}^{ij\rightarrow kl}
\label{2parton}
\end{equation}
with $d\hat\sigma/d\hat t$ containing the squared matrix elements listed in Eqs. (A-1)-(A-8). 
For the parton distribution functions (PDFs), we use the CTEQ6L1 set \cite{CTEQ}. For simplicity in this exploratory work we, however,  neglect the nuclear effects to the PDFs \cite{NPDF,EKS98,EPS09}, since these are small in comparison with the ones arising from the final state interactions with the medium. The initial rapidity \(y_i\) is randomly generated in the range \([y_{min},y_{max}]\) from a flat distribution. (This is consistent on the level of the longitudinal boost-symmetry approximation made for the hydrodynamical part.)
This fixes the hard-parton energy \(E\) and polar angle \(\theta\) of its momentum vector. The azimuth angle \(\phi\) is evenly distributed between \([0,2 \pi]\). 

In order to keep track of the hard partons interactions
with the medium in the coordinate space, the initial spatial position of the hard parton has to be fixed as well. 
We start to follow the hard parton at the initial longitudinal proper time \(\tau_0\) of our hydrodynamical model.
Since in the c.m. frame all hard partons are produced in the Lorentz-contracted overlap region at $z\approx 0$, the longitudinal position at later times (before the first collision at $\tau\ge \tau_0$) is assumed to be determined by the longitudinal momentum only. The initial time and longitudinal coordinates for the hard parton are thus \(t_0 = \tau_0 \cosh{y_i}\) and \(z_0 = \tau_0 \sinh{y_i} \). The initial position in the transverse plane is sampled from the nuclear overlap function
\begin{equation}
\label{taa}
T_{AA}({\bf b})= \int d^2 {\bf s} \, T_A({\bf s}+ {\bf b}/2) T_A({\bf s}-{\bf b}/2),
\end{equation}  
where \({\bf b}\) is the impact parameter. As we focus only on the central collisions, in the following \({\bf b}=0\). The nuclear thickness function \(T_A({\bf s})\) is defined as usual, 
\begin{equation}
\label{nucthick}
T_A({\bf s}) = \int dz \, \rho_A({\bf s},z),
\end{equation}
where for the nuclear density \(\rho_A({\bf r})\) we use the standard Woods-Saxon distribution
\begin{equation}
\label{woodsaxon}
\rho_A({\bf s},z)=\frac{\rho_0}{\exp{\frac{r-R_A}{d}}+1},
\end{equation}
with \(r=\sqrt{|{\bf s}|^2+z^2}\), \(R_A = 1.12A^{1/3}-0.86A^{-1/3}\), \(d=0.54\)~fm and \(\rho_0=0.17\)~fm\(^{-3}\).

The hard parton propagates through the plasma in small time steps \(\Delta t\), during which we propagate the parton in position space. Since we consider only incoherent scatterings here, the probability for not colliding in the time interval \(\Delta t\) is assumed to be given by the Poisson distribution
\begin{equation}
P(\text{No collisions in } \Delta t)=e^{-\Gamma_i \Delta t}, 
\end{equation}
where \(\Gamma_i\) is the total scattering rate \eqref{totgamma} for the hard parton of the type $i$. Hence the probability to collide at least once during the time \(\Delta t\) is \(1-e^{-\Gamma_i \Delta t} \approx \Gamma_i + {\cal O}((\Gamma_i\Delta t)^2)\). For small enough \(\Delta t\) we can assume that there will be at most one collision. 

However, as explained above, we calculate the scattering rates \eqref{totgamma} in the local rest frame of the quark-gluon plasma fluid element and we must boost also the time step \(\Delta t\) to the same frame, 
\[
\Delta t \rightarrow \Delta t' = \frac{\Delta t}{\gamma}, 
\]
where the Lorentz factor \(\gamma = \frac{1}{\sqrt{1-|{\bf u}|^2}}\) appears, \({\bf u}\) being the 3-dimensional flow velocity vector of the hydrodynamical plasma in the c.m. frame of the colliding nuclei. Should a scattering happen, the probability $P_{ij\rightarrow kl}$ for a given type of scattering process is determined by the partial scattering rates \eqref{scattrate} as
\begin{equation}
P(\text{Process } {ij\rightarrow kl}) = \frac{\Gamma_{ij\rightarrow kl}}{\Gamma_i}.
\end{equation} 

We produce the  4-momentum \(p_2\) of the scattering partner from the thermal medium according the energy and angular distributions defined by the differential scattering rate $d\Gamma_i/d^3p_2$ (see Eq.~\eqref{scattrate}) of the process. We then boost the 4-momenta of the hard parton \(p_1\) and the thermal particle \(p_2\) to the c.m. frame of the partonic collision, rotating the frame so that \(p_1\) is on the \(z\)-axis. It is now easy to determine the scattering angle \(\theta^*\) by sampling the distribution 
\[
\begin{split}
\frac{d \sigma}{dt}^{ij\rightarrow kl} &= \frac{|M_{ij\rightarrow kl}|^2}{16\pi s^2} \\
\end{split}
\]
using $\cos \theta^* = \frac{2t}{s}+1$, and sampling the azimuthal angle $\varphi^*$ from a flat distribution.

After determining the scattering angles \(\theta^*\) and $\varphi^*$, we can construct the final state 4-momenta \(p_3\) and \(p_4\) and Lorentz transform these back to the original c.m. frame of the colliding nuclei. The final state parton with highest energy is then chosen as the new hard parton to be propagated further, for which we repeat the procedure outlined above with the next timestep. Note that the type of the hard parton may change here. (We also note that in our code, we have the possibility to follow the recoil partons as well but at sufficiently high parton momenta ($p_T\gsim 5$~GeV), we do not see a significant effect on the observables studied here.) This procedure continues until the surrounding matter has cooled down to the critical temperature \(T_C\). At \(T=T_C\) the system is in this work in a mixed phase where energy density \(\epsilon\) keeps decreasing while the temperature stays constant. 

It is {\em a priori} not clear how the interaction with the medium in the mixed phase should be modelled properly in pQCD. However, we will demonstrate {\em a posteriori} (see Fig. \ref{raapartonQGP} ahead) that within the assumption of an effective temperature, the effect of the mixed phase is small because the medium density is small as compared to the initial density. Thus, unless there is an anomalous enhancement of parton-medium scattering in the mixed phase by non-perturbative processes, a detailed treatment of the mixed phase is not important, and we can also ignore the interactions of the hard parton in the hadron gas. In the following, we treat the interactions in the mixed phase by using an effective temperature
\begin{equation}
\label{efftemp}
T_{eff}=(\frac{30}{g_Q \pi^2}(\epsilon-B))^{1/4},
\end{equation}
where \(g_Q=g_g + \frac{7}{8}2N_fg_q = \frac{95}{2}\) is the quark-gluon plasma degrees of freedom and \(B\) is the bag constant. 
Our simulation ends when \(T < T_C\) and the system enters the hadron gas phase, where we assume no significant interaction between parton and medium.

The outcome of the procedure described above is
a medium-modified distribution of high-energy partons, $\frac{dN^{AA \rightarrow f+X}}{dp_T dy}$. In order to compare with observables, we have to hadronize this distribution. In the current leading-particle case, this can be done 
by convoluting the obtained partonic distribution with the fragmentation function \(D_{f \rightarrow \pi}(z,\mu_F^2)\):
\begin{equation}
\label{hadrondistr}
\begin{split}
\frac{dN^{AA \rightarrow {\pi}+X}}{dP_T dy} = &\sum_f \int dp_T dy \frac{dN^{AA \rightarrow f+X}}{dp_T dy}\cdot\\
&  \int_0^1 dz D_{f \rightarrow \pi} (z,\mu_F^2) \delta(P_T-zp_T),
\end{split}
\end{equation}
where \(z = \frac{P_T}{p_T}\) is the fraction of the final parton momentum \(p_T\) available to the hadron with momentum $P_T$, and \(\mu_F \sim P_T\) is the fragmentation scale. In this work we have used the Kniehl-Kramer-P\"{o}tter (KKP) fragmentation functions \cite{KKP}.

\section{Results}

In the following, we choose a lower bound of the parton transverse momentum, \(p_{T_{min}}\), as 5.0 GeV to account for the observation that at low $P_T$ the main hadron production mechanism is not independent fragmentation.
Partons which fall below $E=3T$ in our simulation
are assumed to be thermalized and become part of the medium, 
and we do not follow these partons anymore. 
We average all observables across the rapidity window $[-0.35,0.35]$ which corresponds to the PHENIX acceptance. 
To make sure we account properly for all the possible partons falling into this final rapidity window, we choose the initial $y_{min}=-y_{max}=-1$. We refer to all partons above both the $p_{T_{min}}$ cut and within the final rapidity acceptance as 'punch-through partons'.

\begin{figure}[h]
\centering
\includegraphics[width=9cm]{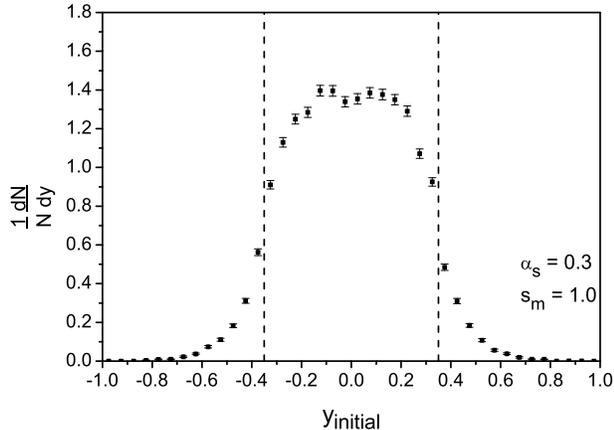}
\caption{The initial rapidity distribution of the punch-through partons. The final interval $|y|\le0.35$ is shown by the dashed lines.}
\label{fig:rapidity}
\end{figure}

To verify the magnitude of the rapidity shifts experienced by the high-$E$ partons, we plot in Fig.~\ref{fig:rapidity} the initial rapidity distribution of the punch-through partons in our default set-up with $\alpha_s=0.3$ and $s_m=1.0$. We can see that about 13 \% of the punch-through partons originate from outside of the final rapidity interval.
On the one hand, this suggests that the eikonal approximation -- where all partons would remain within the interval $|y|\le 0.35$ -- is a fairly good first approximation. On the other hand, if one aims at a precision analysis of the single-hadron distributions or, even more importantly, at correlations of two high-$P_T$ hadrons, the non-eikonal dynamics should  be obviously taken into account. An attempt to quantitatively understand the rapidity shift seen in Fig.~\ref{fig:rapidity} can be found in Appendix~\ref{sec:yshifts}.

Also the leading-parton flavour-conversions are usually neglected in the eikonal energy-loss simulations. Interestingly, this effect is not symmetric between quarks and gluons: due the collision rates, 
there are more flavour conversions related to gluons than to quarks, 
and a quark from a $g\rightarrow q$ leading-parton conversion is then more likely to become a punch-through parton than a gluon from a $q\rightarrow g$ conversion. 
Our simulation shows that for a simplified homogeneous static plasma cylinder of, say, radius $L=5$~fm and $T=500$~MeV, as much as 40 \% of those original high-$p_T$ gluons which will become punch-through partons, will turn into quarks. In the full simulation with 3-d hydrodynamical background medium, however, only about 4\% of all punch-through partons have undergone this effect. We thus conclude that in simulations for single-hadron distributions at RHIC, where a realistically evolving background medium is considered, the leading-parton flavour-conversions can be neglected.  

\begin{figure}[h]
\centering
\includegraphics[width=9cm]{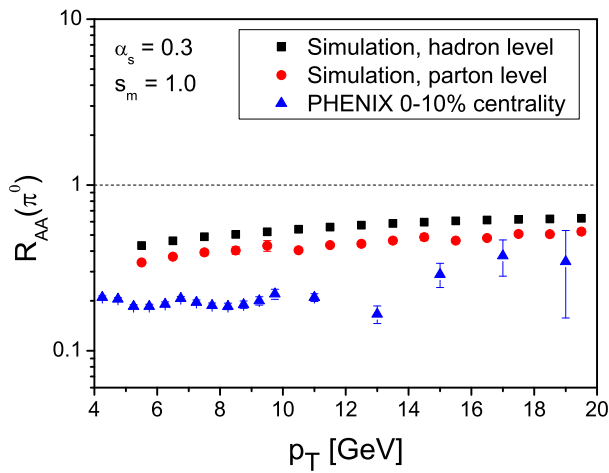}
\caption{(Color online) Comparison of the \(\pi^0\) nuclear modification factor both at hadron and parton level with strong coupling constant \(\alpha_s=0.3\). PHENIX data is from \cite{PHENIX-R_AA}.}
\label{raahadron}
\end{figure}

In Fig.~\ref{raahadron} we present the nuclear modification factor $R_{AA}$,
\begin{equation}
R_{AA}(P_T,y) = \frac{dN^{\pi}_{AA}/dP_Tdy}{T_{AA}(0) d\sigma^{pp}/dP_Tdy}.
\end{equation}
for $\alpha_s=0.3$, $s_m=1.0$, both on the hadronic level and on the partonic level, and compare it with the $\pi^0$ data from PHENIX \cite{PHENIX-R_AA}. It is evident that for this 'default' choice of $\alpha_s$ and $s_m$ elastic collisions are not enough to account for the data but that they should not be neglected, either.
This result is in agreement with the calculation of purely elastic energy loss in the Arnold-Moore-Yaffe (AMY) formalism averaged over hydrodynamics \cite{Ruppert} where a similar value of $R_{AA}\sim 0.6$ is found for $\alpha_s = 0.27$. Note that a comparison with models not using a constrained fluid-dynamics model for the bulk evolution or the same $\alpha_s$ is not meaningful.

It can further be deduced from the figure that since the suppression as a function of $P_T$ (or partonic $p_T$) is flat, the convolution of the medium-modified parton  spectrum with the fragmentation function does not cause a significant effect in $R_{AA}$. We will thus in the following focus on the partonic suppression pattern, with the implication that the partonic suppression provides a reasonable estimate for the suppression also on the level of hadrons. As a technical point, we note that in order to collect enough statistics also at high-$p_T$, the partonic results in Fig.~\ref{raahadron} consist of three dedicated runs, corresponding to \(p_{T_{min}}=5,10,15\)~GeV.

\begin{figure}[h]
\centering
\includegraphics[width=9cm]{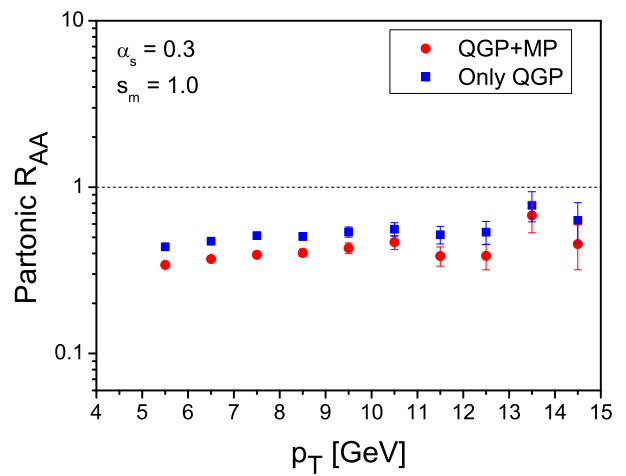}
\caption{(Color online) The parton-level nuclear modification factors with our default set up  \(\alpha_s=0.3\) and $s_m=1.0$, including only the scatterings in the QGP phase (squares) and in the QGP and mixed phase but not the in hadron gas (circles).}
\label{raapartonQGP}
\end{figure}

In Fig.~\ref{raapartonQGP} we demonstrate, as discussed above, that the mixed phase is in this set-up indeed causing only a small additional suppression and, consequently, that it is justified to neglect the hadron gas effects in the present study. The results shown in this and in the following figures are from runs with $p_{T_{min}}=5$~GeV.

\begin{figure}[h]
\centering
\includegraphics[width=9cm]{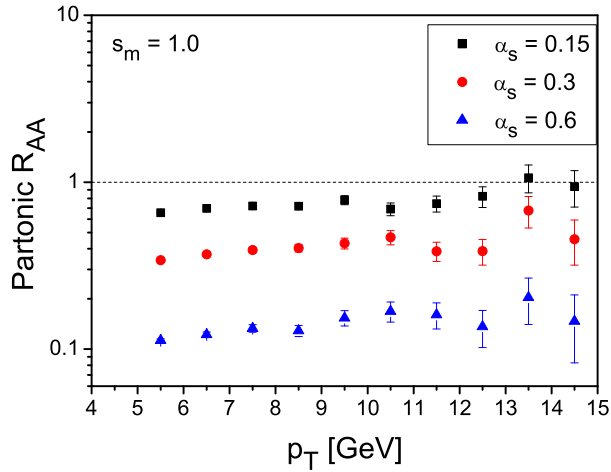}
\caption{(Color online) Parton level nuclear modification factor \(R_{AA}\) for the regulator-mass parameter value \(s_m = 1.0\) and strong coupling constant values \(\alpha_s=0.15,0.3,0.6\).}
\label{raapartonalfa}
\end{figure}

In Fig.~\ref{raapartonalfa} we show the partonic nuclear modification for three different choices of constant \(\alpha_s\). As expected, we observe a rather strong dependence on the value of \(\alpha_s\). The behavior is quite straightforward; doubling the coupling approximately doubles the relative suppression. With \(\alpha_s=0.6\) we find a suppression due to elastic energy loss already below the data. On the other hand, with \(\alpha_s=0.15\) the effect is rather weak. We thus focus on \(\alpha_s=0.3\) in the following when examining other model parameters.

\begin{figure}[h]
\centering
\includegraphics[width=9cm]{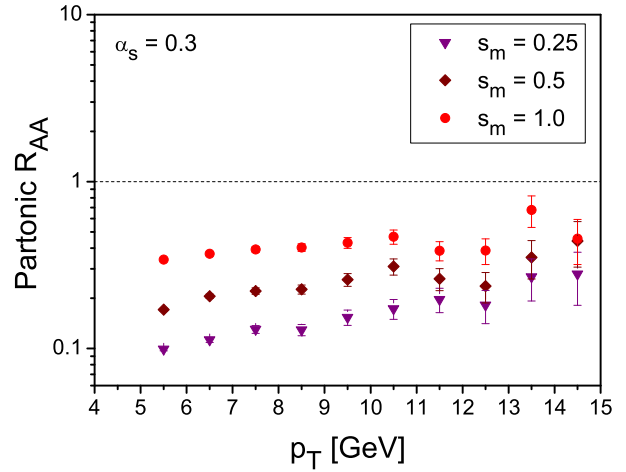}
\caption{(Color online) Parton level \(R_{AA}\) for mass parameter values \(s_m=0.25,0.5,1.0\) and strong coupling constant \(\alpha_s = 0.3\).}
\label{raapartonmass}
\end{figure}

As is obvious the other significant model parameter in our simulation is $s_m$, controlling the 
momentum cutoff scale \(m\) which regulates the scattering cross sections at smallest and largest angles. Like $\alpha_s$, it has a direct effect on the scattering cross sections of a hard parton with medium partons. However, it could be expected to influence energy loss in a different way. While the total cross section can become very large if small-angle forward scattering processes are taken into account, such processes do not transfer a large amount of energy and are hence not efficient for energy loss. Thus, a smaller cutoff mass is expected to increase the total cross section without affecting large-angle scattering whereas a change in $\alpha_s$ affects both forward and large-angle scattering in equal measure. 

A comparison of different values of the cutoff mass in Fig.~\ref{raapartonmass} shows, however, that this parameter has an unexpectedly large effect on the nuclear modification factor. The suppression increases with decreasing mass; this indicates that the cutoff mass is not yet in the region where forward scattering, which is inefficient for energy loss, takes place.

\begin{figure}[t]
\vspace{-0.9cm}

\centering
\includegraphics[width=8.5cm]{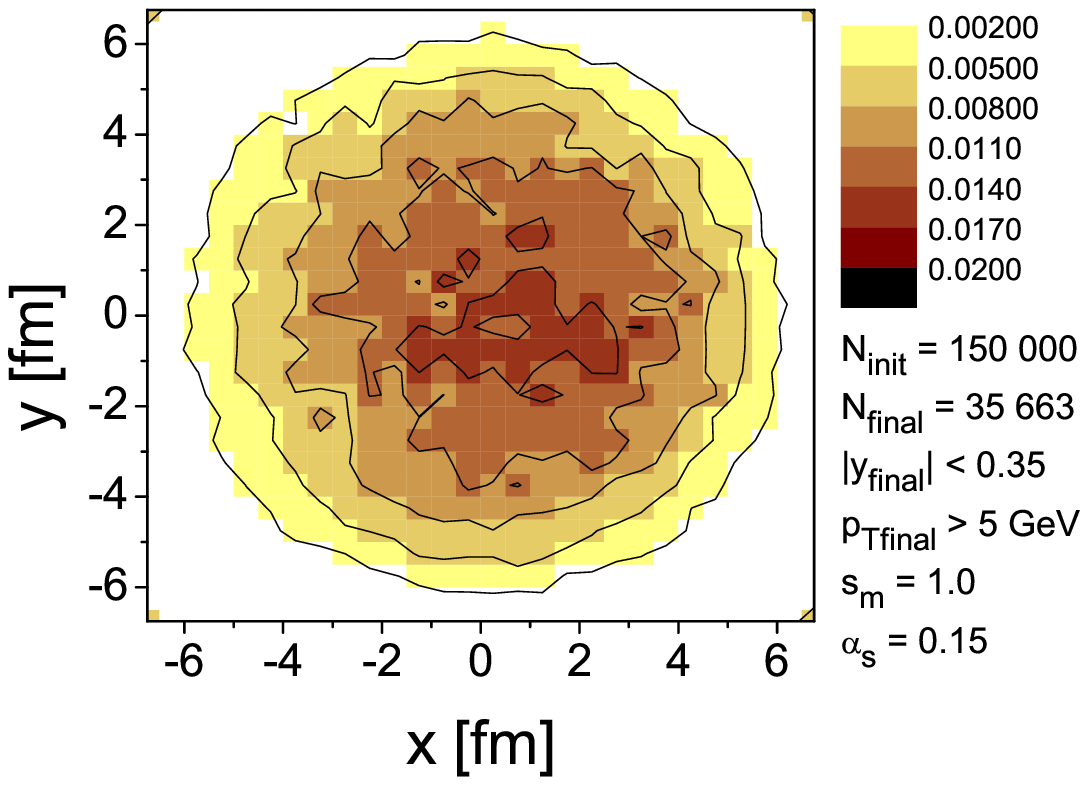}

\vspace{-0.9cm}
\includegraphics[width=8.5cm]{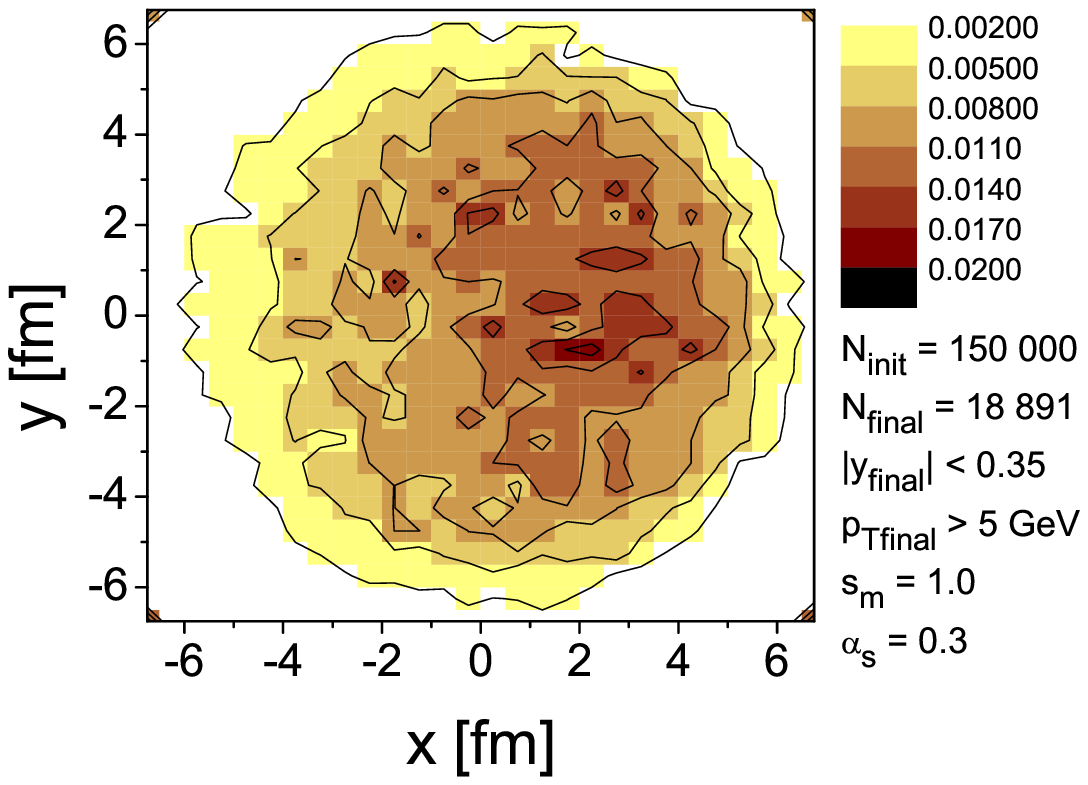}

\vspace{-0.9cm}
\includegraphics[width=8.5cm]{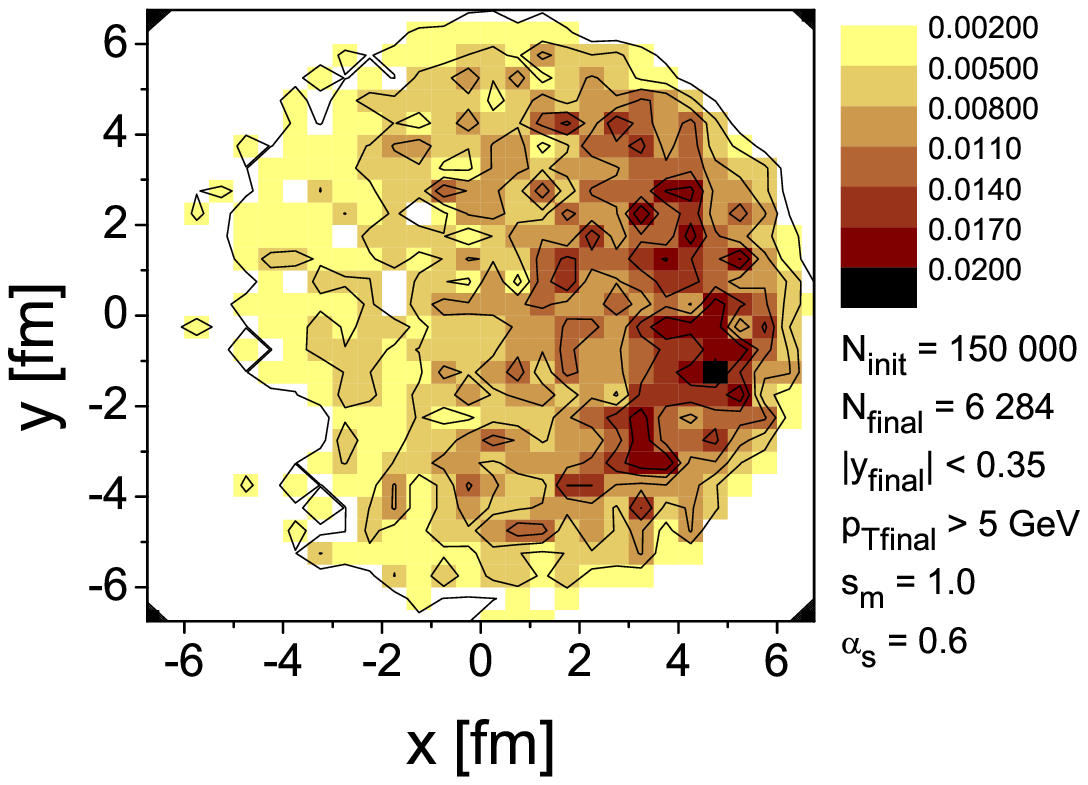}

\caption{(Color online) Initial production points of the punch-through partons on the transverse plane for \(\alpha_s = 0.15\) (above) \(\alpha_s = 0.3\) (middle )and \(\alpha_s = 0.6\) (below). All particles are moving to the positive $x$-direction.}
\label{xyinit}
\end{figure}

In Fig.~\ref{xyinit} we show the geometry underlying the suppression in terms of the initial spatial distribution of punch-through partons for different values of $\alpha_s$. This figure can be considered as a 'tomography' image, where we have rotated the position of each particle in transverse \(xy\)-plane in such a way that they all move to the same direction,  to the direction of positive \(x\)-axis.  With \(\alpha_s = 0.15\) (upper panel), the quark-gluon plasma is quite transparent in transverse direction as far as elastic collisions are concerned, thus the initial distribution of punch-through partons is centered around the origin. Doubling the value of \(\alpha_s\) to 0.3 (middle panel) makes the distribution move off-center. Thus, the medium introduces a surface bias which means that hard partons from the center cannot punch through the matter anymore. Overall, the plasma becomes more opaque, as is expected from Fig.~\ref{raapartonalfa}. The opaqueness is emphasized even more in the bottom panel, which corresponds to $\alpha_s=0.6$,
which reproduces, or slightly overestimates the suppression measured at RHIC. We have also checked the tomography images for different mass parameters ($s_m$) which reproduce the same suppression pattern in $R_{AA}$ as when $\alpha_s$ is varied (see Fig. \ref{raapartonmass}). Our conclusion here is that these do not significantly differ from the ones we have shown in Fig.~\ref{xyinit}.

\section{Discussion}

The results above indicate on the one hand that for our default choice of parameters $\alpha_s = 0.3$ and $s_m = 1$, elastic energy loss is not sufficient to account for the observed value of $R_{AA}$.  On the other hand, we see that at least in the framework which is based on individual parton-parton scatterings instead of, e.g.,  static scattering centers, the elastic energy losses give a non-negligible contribution to $R_{AA}$. This has also been observed by a comparable calculation within the AMY formalism \cite{Ruppert} in which a combination of radiative and elastic energy loss was needed to account for the data. However, our results show an unfortunately strong dependence on cutoff parameters which are in essence supposed to effectively account for the physics of a dynamically running $\alpha_s$ and thermal field-theoretical processes generating a screening mass. Thus for a reasonable variation of these parameters there are pairs $(\alpha_s, s_m)$ for which the $R_{AA}$ data can be described by elastic processes alone. It is thus evident from these findings that a comparison based on the magnitude of the single hadron suppression alone cannot resolve the question how significant the contribution of $2\rightarrow 2$ processes to energy loss actually is. To improve upon this problem, the next step is to include $2\rightarrow 3$ processes and coherence effects into the simulation. 

However, a comparison of the tomographic pictures in this paper (Fig.~\ref{xyinit}) with the results of a radiative energy loss calculation \cite{Correlations2} indicates that the way hard partons probe the medium geometry through energy loss is quite different for radiative and elastic energy loss. Thus, as outlined in a schematical way in \cite{ElasticPhenomenology} it can be expected that pathlength-dependent observables, like the suppression as a function of the angle of the observed hadron with the reaction plane, or the suppression of back-to-back correlations, will show pronounced differences between elastic and radiative energy loss. The MC code introduced here provides an excellent tool to explore this idea, and this will be addressed in a future publication. 

\appendix

\section{Analytical expressions for scattering rates}
\label{scaratapp}

\renewcommand{\theequation}{A-\arabic{equation}}
\setcounter{equation}{0}
The spin- and color summed/averaged squared parton scattering amplitudes $|M|_{ij \rightarrow kl}^2$ for gluons and quarks are (see e.g. \cite{SCP})
\begin{eqnarray}
\begin{array}{c}
ij \rightarrow kl 
\end{array}
& |M|_{ij \rightarrow kl}^2 \nonumber \\
\hline
\label{ggggsirm}
\begin{array}{c}
gg \rightarrow gg 
\end{array}
& \quad \frac{9}{2} g_s^4 \left( 3-\frac{ut}{s^2}-\frac{us}{t^2}-\frac{st}{u^2} \right) \\
\label{ggqqsirm}
\begin{array}{c}
gg \rightarrow q \bar{q} 
\end{array}
& \quad  \frac{3}{8} g_s^4 \left( \frac{4}{9} \frac{t^2+u^2}{tu}-\frac{t^2+u^2}{s^2} \right) \\
\label{gqgqsirm}
\begin{array}{c}
gq\rightarrow gq \\
g\bar q\rightarrow g\bar q
\end{array}
& \quad g_s^4 \left( \frac{s^2+u^2}{t^2}-\frac{4}{9} \frac{s^2+u^2}{su} \right) \\
\label{qiqjsirm}
\begin{array}{c}
q_i q_j\rightarrow q_iq_j \\
q_i \bar q_j\rightarrow q_i \bar q_j \\
\bar q_i q_j\rightarrow \bar q_i q_j \\
\bar q_i \bar q_j\rightarrow \bar q_i \bar q_j
\end{array}
& 
{\quad \frac{4}{9} g_s^4 \frac{s^2+u^2}{t^2}, \quad i\ne j}{~}\\
\label{qiqisirm}
\begin{array}{c}
q_i q_i\rightarrow q_i q_i \\
\bar q_i \bar q_i\rightarrow \bar q_i \bar q_i
\end{array}
&  \quad \frac{4}{9} g_s^4 \left(\frac{s^2+u^2}{t^2}+\frac{s^2+t^2}{u^2}-\frac{2}{3} \frac{s^2}{tu} \right) \\
\label{qaqijsirm}
\begin{array}{c}
q_i \bar{q}_i \rightarrow q_j \bar{q}_j 
\end{array}
& \quad \frac{4}{9} g_s^4 \frac{t^2+u^2}{s^2} \\
\label{qaqiisirm}
\begin{array}{c}
q_i \bar{q}_i \rightarrow q_i \bar{q}_i 
\end{array}
& \quad  \frac{4}{9} g_s^4 \left( \frac{s^2+u^2}{t^2}+\frac{t^2+u^2}{s^2}-\frac{2}{3} \frac{u^2}{st} \right) \\
\label{qaqggsirm}
\begin{array}{c}
q \bar{q} \rightarrow gg 
\end{array}
& \quad  \frac{8}{3} g_s^4 \left( \frac{4}{9} \frac{t^2+u^2}{tu}-\frac{t^2+u^2}{s^2} \right)
\end{eqnarray}
Inserting the expression \eqref{gencrosssection} for the cross section into the definition of the \(\Omega\)-function \eqref{omegafunction} we have
\begin{equation}
\label{omega_amplitude}
\begin{split}
\Omega_{ij\rightarrow kl}&(E_1,E_2,m^2) \\
&=\frac{1}{1+\delta_{kl}} \frac{1}{16 \pi} \int_{2m^2}^{4E_1E_2}ds \left[ \frac{1}{s} \int_{-s+m^2}^{-m^2} dt |M|_{ij\rightarrow kl}^2 \right].
\end{split}
\end{equation}
Several amplitudes contain identical terms. We list the integrals for each of these terms separately, as it is then easy to combine the appropriate terms to form the final expressions for \(\Omega_{ij\rightarrow kl}(E_1,E_2,m^2)\). 
\begin{widetext}
\begin{eqnarray}
\begin{array}{c}
X(s,t,u)
\end{array}
& \int_{2m^2}^{4E_1E_2}ds \left[ \frac{1}{s} \int_{-s+m^2}^{-m^2} dt X(s,t,u) \right] \nonumber \\
\hline
\begin{array}{c}
1 
\end{array}
& 4E_1 E_2 -2m^2 \log{\frac{2E_1 E_2}{m^2}}-2m^2 \quad \\
\begin{array}{c}
\frac{ut}{s^2}
\end{array}
& \frac{2}{3} E_1 E_2-\frac{3}{4}m^2+\frac{m^4}{4E_1 E_2}-\frac{m^6}{48E_1^2 E_2^2} \quad \\
\begin{array}{c}
\frac{su}{t^2},\frac{st}{u^2}
\end{array}
&  -\frac{8E_1^2 E_2^2}{m^2} + 4E_1 E_2 \log{\left(\frac{4E_1 E_2}{m^2}-1 \right)} 
+2m^2 \quad \\
\begin{array}{c}
\frac{t^2+u^2}{s^2}
\end{array}
&  \frac{8}{3} E_1 E_2 -2m^2 \log{\frac{2E_1 E_2}{m^2}} 
 -\frac{1}{2}m^2-\frac{m^4}{2E_1 E_2}+\frac{m^6}{24E_1^2 E_2^2} \quad \\
\begin{array}{c}
\frac{t^2+u^2}{tu}
\end{array}
&  2(4E_1 E_2-m^2) \log{\left(\frac{4E_1 E_2}{m^2} -1 \right)} - 16E_1 E_2  
	+4m^2\log{\frac{2E_1 E_2}{m^2}}+8m^2 \quad \\
\begin{array}{c}
\frac{s^2+u^2}{t^2},\frac{s^2+t^2}{u^2}
\end{array}
&  \frac{16E_1^2E_2^2}{m^2}-8E_1 E_2 \log{\left(\frac{4E_1 E_2}{m^2}-1 \right)} +4E_1 E_2
 -2m^2 \log{\frac{2E_1 E_2}{m^2}}-6m^2 \quad \\
\begin{array}{c}
\frac{s^2+u^2}{su}
\end{array}
&  
 -(4E_1 E_2-m^2) \log{\left(\frac{4E_1 E_2}{m^2}-1\right)} +2E_1 E_2 
 +m^2 \log{\frac{2E_1 E_2}{m^2}}-m^2 \quad \\
\begin{array}{c}
\frac{s^2}{tu}
\end{array}
& 2(4E_1 E_2-m^2) \log{\left(\frac{4E_1 E_2}{m^2}-1\right)} -8E_1 E_2  +4m^2 \quad \\
\begin{array}{c}
\frac{u^2}{st}
\end{array}
& -(4E_1 E_2-m^2) \log{\left(\frac{4E_1 E_2}{m^2}-1\right)}
+10E_1 E_2 -3m^2\log{\frac{2E_1 E_2}{m^2}}-5m^2 
  \quad 
\end{eqnarray}
\end{widetext}

\section{Rapidity shift estimates}
\label{sec:yshifts}
\renewcommand{\theequation}{B-\arabic{equation}}
Here we try to understand quantitatively the magnitude of the rapidity shifts of punch-through partons shown in Fig.~\ref{fig:rapidity}. At the high-energy limit  $E_1\gg T, m$, a straightforward calculation in the static plasma approximation (see Sec.~\ref{subsec:rates}) gives  the average scattering angle in the CMS frame as
\begin{equation}
\langle \cos \theta_3^*\rangle = 1+\langle 2t/s\rangle \approx 1 - \frac{m^2}{2E_1T}\log(\frac{4E_1T}{m^2}).
\end{equation}
We set the initial high-$E$ parton at zero rapidity and its momentum ${\bf p_1}$ to point into the direction of the $x$-axis, while the beam direction is, as usual, the $z$-axis. Having the maximum rapidity shift in mind, we take the medium-parton's momentum ${\bf p_2}$ to reside in the $(x,z)$-plane, its energy to be $E_2=3T$ and the collision (polar) angle relative to ${\bf p_1}$ such that  $\cos\theta_{12}=-1/3$, which correspond to the averages in Eq.~\ref{averages}. Thus 
\begin{equation}
{\bf p_1} = {\bf \hat e_x} E_1 , \quad\quad {\bf p_2} = {\bf \hat e_x} E_2 \cos\theta_{12} + {\bf \hat e_z} E_2 \sin\theta_{12}. 
\end{equation}

In a reverse boost from the CMS frame of the parton-parton scattering, the energy and momentum of the scattered high-E parton become
\begin{eqnarray}
E_3 &=& \gamma(E^* + {\bf v}_{\rm CMS}\cdot {\bf p_3}^*) \\
{\bf p_3} &=&  {\bf p_3}^* + \gamma {\bf v}_{\rm CMS} (\frac{\gamma}{\gamma+1}{\bf p_3}^*\cdot {\bf v}_{\rm CMS} + E^*), 
\end{eqnarray}
where the velocity of the CMS frame is ${\bf v}_{\rm CMS}= ({\bf p_1}+{\bf p_2})/(E_1+E_2)$, $\gamma$ is the Lorentz gamma factor and the asterisks denote the CMS-frame quantities. 
The momentum of the scattered high-$E$ parton in the CMS-frame can be computed to be 
\begin{eqnarray}
{\bf p_3}^* &=& {\bf \hat e_x}[-p_{1z}^*\sin\theta_3^*\sin\phi_3^* + p_{1x}^*\cos\theta_3^*] \nonumber\\
&+& {\bf \hat e_y}E^*\sin\theta_3^*\cos\phi_3^* \\
&+& {\bf \hat e_z}[p_{1x}^*\sin\theta_3^*\sin\phi_3^* + p_{1z}^*\cos\theta_3^*],  \nonumber
\end{eqnarray}
where $\theta_3^*$ is the scattering (polar) angle relative to ${\bf p_1}^*$ and $\phi_3^*$ is the azimuthal angle around 
${\bf p_1}^*$, measured from the $y$-axis, and where the CMS-frame quantities are obtained in a boost to ${\bf v}_{\rm CMS}$ as
\begin{eqnarray}
E^* &=& \gamma(E_1 - {\bf v}_{\rm CMS}\cdot {\bf p_1}) \\
{\bf p_1}^* &=&  {\bf p_1} + \gamma {\bf v}_{\rm CMS} (\frac{\gamma}{\gamma+1}{\bf p_1}\cdot {\bf v}_{\rm CMS} - E_1).
\end{eqnarray}

The point here is that although in the CMS frame (where ${\bf p_2}^*=-{\bf p_1}^*$) the scattering cross section does not depend on the angle $\phi^*$, a $\phi^*$ dependence is generated in the boost back to the plasma rest frame. The rapidity shift $\delta y = y_3$ is now easy to obtain numerically from the components $E_3$ and $p_{3z}$ and we find that to a good approximation 
\begin{equation}
\delta y \approx A\sin\phi_3^*. 
\label{deltay}
\end{equation} 
For a hard parton of, say,  $E_1=10$~GeV in a $T=500$~MeV plasma, we obtain $A\approx 0.26$. The maximum shifts $\delta y \approx \pm A$ occur at $\phi^*=\pi/2$ and $3\pi/2$, i.e. when the scattering remains in the ($x,z$)-plane. Similarly, we find that $E_3/{\rm GeV}\approx 8.7 + 1.4\sin\phi_3^*$ and  $\theta_3\approx 15^{\circ} + 1.1^{\circ}\sin\phi_3^*$.

To estimate the maximum possible total rapidity shift $|\Delta Y|$ induced by $n$ scatterings, we may assume that 
at the limit of  small scattering angles (small rapidity shifts) the maximum-shift distribution $\delta y \approx A\sin\phi^*$ holds in each scattering. The rapidity shifts are additive and we can write the probability distribution for $\Delta Y$ as follows:
\begin{equation}
P(\Delta Y) = \int d\delta y_1\dots d\delta y_n P(\delta y_1)\dots P(\delta y_n)\delta(\Delta Y-\sum_i^n \delta y_i),  
\end{equation}
where the probability distribution for a rapidity shift in one collision, 
\begin{equation}
P(\delta y) = \frac{1}{\pi\sqrt{A^2-(\delta y)^2}}, 
\end{equation}
where $|\delta y|\le A$,
is obtained from the corresponding flat azimuthal angle distribution $P(\phi^*)=1/(2\pi)$ by a change of variables according to Eq.~\ref{deltay}. Now it is easy to obtain
\begin{equation}
\langle (\Delta Y)^2\rangle = n\langle(\delta y)^2\rangle = n\frac{A^2}{2}. 
\end{equation}

For an order of magnitude estimate, we take $n\approx 6$ (from our simulation, since for this static high-$T$ plasma test case the leading-parton flavour-coversions are in fact a non-negligible effect), and  arrive at the average maximum total rapidity shift 
$|\Delta Y|=\sqrt{\langle (\Delta Y)^2\rangle}= A\sqrt{n/2}\approx 0.45$. 

\begin{figure}[t]
\centering
\includegraphics[width=9cm]{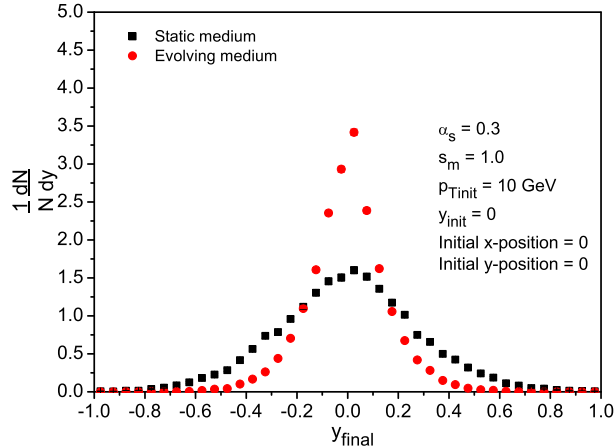}
\caption{(Color online) Rapidity spread of punch-through partons which originate from a 10 GeV, $y=0$ parton at the spatial origin.}
\label{fig:yshifts}
\end{figure}

Figure~\ref{fig:yshifts} shows our simulation test results for the rapidity spread of punch-through partons which originate from
10 GeV partons sent off from the origin into the $x$-direction. 
The width of the distribution in the static 500 MeV plasma case (squares) is seen to be about 0.3, which nicely confirms the order of magnitude of our upper limit estimate $|\Delta Y|$. The smaller width in the simulation follows from the fact that the simulation accounts for scatterings (corresponding to a continuum of azimuthal incoming angles $\phi_{12}$ around ${\bf p_1}$), where the rapidity shift effects are smaller than in our estimate above. Changing the background into a hydrodynamically evolving more realistic medium (circles) reduces the width of the rapidity shift by about a factor 3. This is caused by the fact that collision rates, scattering angles $\theta_3^*$ and rapidity kicks all decrease with decreasing temperature.
Finally, in a full simulation, where the high-$E$ partons can be produced anywhere in the transverse plane, and where also the surface bias effect is present (see Fig.~\ref{xyinit}), the rapidity shifts become still somewhat smaller. From these considerations, and realizing that the rapidity spread distributions should be applied to each final-rapidity bin of Fig.~\ref{fig:rapidity}, we conclude that the degree of non-eikonality ($y$-shifts) is understood, and also that if we wish to fully account for punch-through partons in the finite rapidity interval $|y|\le 0.35$, we have to consider an initial window of $|y_i|\lsim 1$.

\begin{acknowledgments}
We thank Harri Niemi for providing us with the hydrodynamic simulations.
J.A. gratefully acknowledges the grant from the Jenny and Antti Wihuri Foundation.
This work was also financially supported by the Academy of Finland: 
K.J.E.'s Projects 115262 and 133005, and T.R.'s Academy Research Fellowship with the associated Project 130472.
\end{acknowledgments}

\end{document}